# The EU AI Act in Development Practice: A Pro-justice Approach

Tomasz Hollanek Leverhulme Centre for the Future of Intelligence, University of Cambridge, th536@cam.ac.uk

Yulu Pi University of Warwick and University of Cambridge, yulu.pi@warwick.ac.uk

Dorian Peters Imperial College London and University of Cambridge, d.peters@imperial.ac.uk

Selen Yakar University of Cologne, s45syaka@uni-bonn.de

Eleanor Drage Leverhulme Centre for the Future of Intelligence, University of Cambridge, ed575@cam.ac.uk

Abstract

With the AI Act's adoption in the European Union, there is a pressing need for companies obligated to comply with the Act to grasp and implement the necessary compliance requirements. However, compliance can be challenging, especially when navigating ambiguous aspects of the Act, such as risk management or fundamental rights impact assessment. In this paper, we explain how a comprehensive toolkit that we have developed to help companies meet this challenge, the High-risk EU AI Act Toolkit (HEAT), puts forward a pro-justice, feminist ethics-informed approach to enhancing compliance with the EU AI Act. While HEAT helps development teams meet the Act's requirements, here we explain how feminist and other pro-justice ethical frameworks allow for meaningful translation of AI ethics theory into practice, moving beyond mere compliance with emerging regulation. We explain how the pro-justice approach resulted in our development of specific methods and guidelines included in HEAT: we explain how the feminist understanding of expertise and the acknowledgment of partiality of knowledge has shaped the HEAT method for engaging different stakeholders, including marginalized and minoritized groups; we illustrate how our pro-justice approach expands on the Act's minimal requirements in relation to accessibility through the incorporation of disability justice and design justice guidance; we expound on how we encourage HEAT users to engage with feminist data science and how our pro-justice interpretation of the Act presents environmental concerns as inseparable from other concerns about the social impact of AI. Like pro-justice theories, the Act is utopian and aspirational, calling for and demanding a fairer world. The theories we draw on are not naively utopian. Instead, they inspire non-innocent approaches to interrogating normativity in systems of all kinds - technological and otherwise. By this we mean that pro-justice orientations are cognizant of their involvement in the systems they seek to critique, and offer best practices with how to grapple with the trade-offs inherent in reducing and eradicating harmful behaviour from within. These best practices, as we explain in this paper, are what HEAT both embodies and enables.

1. Introduction

With the European Union's AI Act officially entering into force in August 2024, there is a pressing need for companies obligated to comply with the Act to grasp and implement the necessary compliance requirements. Mitigating the risks associated with AI systems is not just a moral imperative but a legal obligation. Taking a risk-based approach, the Act categorises AI systems into different risk tiers based on their potential risks and level of impact, each with its own set of rules to ensure safety and responsibility.

For high-risk AI systems, the Act imposes rigorous requirements covering the entire lifecycle, including data governance, model's accuracy, robustness, cybersecurity, and post-market monitoring. However, compliance can be challenging, especially when navigating ambiguous aspects of the Act, such as risk management (Article 9 of the AI Act [14]). Additionally, compliance is complicated by the fact that AI systems are developed and deployed by a diverse array of actors - including data scientists, software engineers, user experience designers, and product managers (PMs) to name just a few. Each of these actors brings distinct disciplinary expertise, practices, and organizational responsibilities. [33, 51]. To ensure compliance, organizations must interpret and translate the Act's requirements into actionable tasks tailored to these varied roles to achieve compliance.

The pressure for compliance, coupled with a heightened understanding of AI's potential negative social impact and the necessity to develop responsible AI responsibly, has led to the proliferation of AI ethics compliance toolkits, tools, and guidelines [28]; The introduction of the AI Act is bound to result in the development of new tools meant to assist AI development teams in comprehending, fulfilling, and documenting their efforts to adhere to the AI Act. [33, 51]. Critical of what is already available within the AI ethics toolkit 'landscape' and mindful of what is still missing, in this paper we explain how a toolkit that we have developed, the High-risk EU AI Act Toolkit (HEAT), puts forward a pro-justice approach to enhancing compliance with the EU AI Act. HEAT aims to help high-risk AI providers meet and exceed the Act's requirements by offering an interpretative and compliance framework grounded in intersectional feminism and other pro-justice ethics. We chose these perspectives to fully capture the spirit and intent of the Act since, as we explain in what follows, the ethical and responsible development and deployment of AI is only possible when we attend to a wider context within which AI systems are shaped and continuously shaping. This pro-justice approach is deeply embedded in both the structural design of the HEAT toolkit and the tasks it presents for users to study and complete. HEAT moved away from the checklist format and towards interconnected spaces that require continuous engagement from different team members. This design encourages ongoing and iterative efforts to AI Act compliance, where users are prompted to revisit and reconsider areas of concerns throughout the AI system's life cycle. The tasks within the toolkit are carefully crafted to reflect the high risk AI obligations from the Act while focusing on power dynamics, structural injustices, and the broader societal context with real-world examples.

This paper examines the pro-justice theories underpinning the toolkit and provides concrete examples to illustrate our approach. The toolkit was developed in collaboration with an industry partner and an advisory board comprising researchers, industry experts, and policy practitioners from various disciplines and global contexts. Detailed discussions of the design process and user testing will be addressed in subsequent publications. This paper draws on interdisciplinary research in human-computer interaction, critical design, feminist studies, and pro-justice theories, as outlined in section 2, 'Background and Related Work'. Section 3 describes how we applied feminist and pro-justice methods to achieve a practical and real-world interpretation of the EU AI Act. Section 4 offers specific examples of how these approaches were integrated into the design and content of the toolkit.

## 2. Background and Related Work

2.1 On the 'toolkittifcation' of AI ethics and the need for feminist and pro-justice approaches to practice.

The growing awareness of AI's potential for harm and its real negative social impacts, combined with the pronounced need for practical and actionable guidelines for ethical and responsible AI development, have led to a proliferation of AI ethics toolkits [28]. There is no single and consistent definition of a design toolkit [39] and, given the variety of available resources, an AI ethics toolkit can be defined broadly as 'any type of document, product, or website whose explicit aim is to facilitate change in the ways in which AI systems are designed, deployed, or governed and to ensure that the results of development lead to desirable social consequences' [28]. The 'toolkittification' of ethics work for AI development and governance has been the subject of attempts to map the 'landscape' of AI ethics toolkits, pointing to the similarities in approaches adopted by different toolkit makers and the related pitfalls in the design of the toolkits themselves [3, 4, 24, 34, 37, 41, 53]. A recent meta-review of AI ethics toolkits points to several issues inherent to the toolkit format: the gaps in the lists of values and principles that the toolkits are meant to help operationalize (with 'solidarity' or 'environmental sustainability' consistently omitted); the oversimplified, techno-solutionist understanding of ethics that informs the toolkits' design; and the co-optation of concepts such as participatory or inclusive design decontextualized in ways that foreclose their meaningful adoption in AI development [28].

Despite this and other critiques of the toolkit format in AI ethics work [26, 40], the adoption of the the European Union's AI Act [14] – the first comprehensive piece of legislation meant to ensure the development of the technology is carried out responsibility – is resulting in the production of new tools to assist AI development teams in understanding, fulfilling, and documenting their efforts to adhere to the new regulation. Indeed, such additional assistance is needed as one of the Act's requirements for producers of high-risk AI systems is to conduct an assessment of the system's impact on fundamental rights and values. This is a challenging task for philosophers, let alone software engineers or product managers.

Landscape studies of AI ethics toolkits point to several, consistent recommendations for future toolkit makers. The toolkit makers are encouraged to:
- integrate diverse ethical perspectives, such as feminist and indigenous viewpoints when designing toolkits;
- position ethics work as a matter of collective deliberation and action, both within and outside of a specific development team, and affecting not only design but also business decision-making;
- give structure to ethical deliberation in design while acknowledging the inherent complexity of ethical challenges, paying attention to both direct and indirect impacts of AI systems; allow for the 'ongoingness' of ethical practice and reject a box-ticking logic that many toolkits embrace; and
- foreground power-sensitive approaches to participatory design as integral to responsible AI development [28, 53].

Recognizing the genuine need for new resources that support compliance with the Act, and bearing in mind the existing critiques of the AI ethics toolkits' landscape including the recommendations for future toolkits summarised above, in what follows we expound on our approach to applying these recommendations to translating AI ethics 'theory' into 'practice' through HEAT. This approach, as we already mentioned, is grounded in feminist, anti-racist, and pro-justice critiques of technology. These critiques recognize that ethical development is only possible when we attend to the wider contexts within which AI technologies function, perceive and challenge the power dynamics that determine who will benefit and who will bear the brunt of technological development, and understand structural injustices that result in persistent harms, perpetuated by new technological solutions. HEAT is intended to help developers comply with the Act, but by adopting the pro-justice lens to interpreting the Act, we also enable moving beyond mere compliance – as the necessary step in building new technologies ethically and responsibly.

It is also important to note at the outset that our approach to the development of HEAT has also been informed by the critiques of existing toolkits that focus on their usability and discoverability. Despite the growing number of AI ethics toolkits, studies show that these resources are seldom used by industry practitioners [43, 49]. Researchers attribute this disinterest to several factors: many toolkits are developed in academic settings without collaboration with industry professionals, leading to a mismatch with practitioners' needs [43, 49]; the user experience is often cumbersome, with some toolkits causing information overload [34]; many practitioners are unaware of the wide range of available toolkits due to disciplinary silos and ineffective promotion strategies targeting the intended audience [39]. We should stress that, to avoid these pitfalls, we developed HEAT in collaboration with industry partners and the workflow we propose in the Toolkit has been iteratively co-designed with product managers and machine learning practitioners. A detailed discussion of how we addressed direct usability concerns and conducted user testing of HEAT – and how this influenced our design choices – is, however, beyond the scope of this paper (and will be discussed elsewhere), as here we focus specifically on how HEAT puts forward a feminist theory-informed interpretation of the Act.

2.2 Feminist, antiracist and projustice interpretations of the EU AI Act

HEAT aims to meet and exceed the Act's requirements by providing an interpretative framework with which to understand and act on its instructions. Our project began by asking: which animating theory from AI ethics and beyond can best achieve this goal? We chose intersectional feminism and other pro-justice ethics that address race, gender, class, disability, age and environmental sustainability, as we believe they animate the spirit as well as the letter of the law. Representative democracy is foundational to the EU, making our pro-justice emphasis on accessibility, inclusivity, and politics also particularly meaningful. Like pro-justice theories, the Act is utopian and aspirational, calling for and demanding a fairer world. The theories we draw on are not naively utopian. Instead, they inspire non-innocent approaches to interrogating normativity in systems of all kinds - technological and otherwise. By this we mean that pro-justice orientations are cognizant of their involvement in the systems they seek to critique, and offer best practices with how to grapple with the trade-offs inherent in reducing and eradicating harmful behaviour from within.

Pro-justice theories and methodologies do not deal in abstract thought experiments, instead gaining knowledge through real-world, context-based use cases and non-expert knowledge. For example, we use feminist theory to demonstrate how power informs and shapes our material lives, including the unequal power dynamics between the company and the consumer (which pre-exist the product ideation stage). This informs our translation of key concepts from the EU Act ('redress', 'harm' 'bias' 'democracy' 'consent' 'complaint') from the abstract to the particular in ways that previous studies have shown are meaningful to practitioners [10]. In doing so, we go beyond mere legal compliance by prompting product managers to think critically about how AI relates to structural inequality and engage meaningfully with the ethos of the Act. Below we focus on the three key approaches we took to translate pro-justice principles to actions for product managers.

3. Applying our Pro-Justice Lens

3.1 Making theory actionable

Our task was also to use pro-justice ideas to express the Act's demands as clearly, concretely and practically as possible, bearing in mind that the law is only as good as its implementation. We did this through practical instructions, questions and tasks that go beyond impact assessments and checklists (which often only flag concerns after harm has already been inflicted) by allowing companies to be proactive about building good technology. For example, in demonstrating how power informs and shapes our material lives in relation to key concepts from the Act (i.e. 'redress', 'harm' 'bias' 'democracy' 'consent' 'complaint'), feminist theory enriches PMs' understanding of how harmful effects arise as a result of the unequal power dynamics between the company and the consumer. We inform and upskill PM's on these issues through learning devices like short videos that help make the Act's obligations more meaningful to PMs.

3.2 A socio-technical approach

One of the primary successes of feminist science studies is breaking down the barrier between technology and culture [6, 25, 52]. AI systems are inherently sociotechnical as their technical components are embedded in social environments, resulting in significant impacts on individuals and society as a whole. Therefore, ensuring compliance with the Act requires consideration of overlapping technical, social and legal dimensions. While we do provide mechanisms for PMs to clearly explain technical choices and their implications, ensuring accountability and fostering trust with stakeholders, we also set out to explore the limits of both technical and sociotechnical approaches to AI ethics. When are fairness mechanisms useful? When are they not? How can humanities approaches be reshaped to speak to people with technical backgrounds? Every approach has its limitations, therefore we encourage practitioners to embrace intellectual humility by gaining an awareness of other approaches and embracing collaborations to address issues holistically. Beyond mere documentation, we guide projects in understanding, choosing and reasoning key design choices from socio-techno-environmental perspectives. This tool therefore mobilises both computer science and humanities expertise to experiment with combining the best technical methods (from our own experts, as well as from external stakeholders like Hugging Face and DAIR), with a deeper, socio technical understanding of how AI relates to structural inequality more broadly - and therefore how systems can be harmful even when error or bias-free.

3.3 From bias to power

Often, the term bias does not communicate the full extent of a harm created by an AI product. Issues not often captured by the term include the spread of disinformation, decreasing social cohesion and trust, and the exaggeration or miscommunication of a product's capabilities (e.g. AI hiring tools that claim to strip race and gender from a candidate's profile). Bias' numerous mathematical definitions also make the term ambiguous at best and unworkable at worst for engineers [10]. Lastly, approaches to 'debiasing' often rest on the assumption that technology is essentially neutral and devoid of politics and can, when debiased, perform objectively. Instead we draw on the maxim "don't ask if artificial intelligence is good or fair, ask how it shifts power" [32] by informing PM's about how harms arise from power imbalances. We recommend solutions such as 'algorithmic reparations' to shift power in the direction of users, and to expand and shake up internal conversations about what constitutes 'bias work' [18]. Thinking with power, rather than bias (i.e. the complex and interlinked oppressions people face as subjects of power, rather than distinct forms of bias 'gender bias'), allows us to approach pro-justice thinking as intersectional. We also encourage internal conversations about what companies are willing to do - or not do - as a way of shifting power. We propose that all AI is political, and therefore companies should be aware of the politics of their products.

3.4 Collective deliberation

Following [47], we aim to redistribute expertise by shifting the idea of the 'AI ethics expert' from an individual person to collective expertise, which various stakeholders possess only in partiality. The tool's participatory design and stakeholder engagement mechanisms prompt companies to tap into expertise through collaborations from the product ideation stage onwards. While there is now lots of work on 'stakeholder-first' approaches, what this entails is rarely defined, potentially leading to 'participation-washing' [46] We take a power-sensitive approach and give clear instructions on how to involve different stakeholder groups – and minoritized groups in particular – at different stages of the development and deployment. We should note that the act of collective deliberation and the idea of distributed expertise have also informed the design of HEAT itself. The full design process (and meta-design) of the toolkit itself will be explored in a subsequent paper; here, we focus specifically on how HEAT guides its users through co-design and stakeholder participation as fundamental to a pro-justice interpretation of the Act.

4. A Pro-Justice Approach to the Act in Practice

Through HEAT we challenge the logic of the checklist as the primary instrument of AI ethics – a set of tasks to be completed by different teams in a predetermined sequence – and we move towards the logic of spaces: interconnected areas of concern that different team members must engage with and continue revisiting throughout software development and deployment. HEAT comprises seven of such spaces:

> Space 1. Business, team, principles: Here, HEAT users are prompted to consider the makeup of their own teams, the values that guide development, to critically analyse the problem the team wants to solve, and to justify the use of AI.
>
> Space 2. Impact, risk, mitigation: In this space, teams identify potential harms related to the development of a given system and formulate a risk management strategy.
>
> Space 3. Stakeholder engagement: This space guides HEAT users through the process of consulting and including different stakeholders in the design process.
>
> Space 4. User-centred design: Here, teams are prompted to consider the user-facing elements of the system, to ensure accessibility and transparency (considering disability and design justice principles), as well as  meaningful consent from users and enabling them to seek redress if they are negatively affected by a given system.

Space 5. Data governance: In this space, teams can document their approach to data collection and processing, including GDPR compliance, and learn about data justice.

Space 6. Model governance: This space guides teams through the process of documenting their approach to model training and selecting appropriate fairness metrics.

Space 7. Evaluation and care: In this space, teams are instructed not only to monitor a given system after deployment but also establish effective procedures for responding to reported incidents and complaints.

The diagram below (Figure 1.) summarises the workflow proposed by HEAT. In this section, we provide several examples of how the feminist lens allowed us to interpret the Act in the construction of these spaces and individual tasks that HEAT comprises.

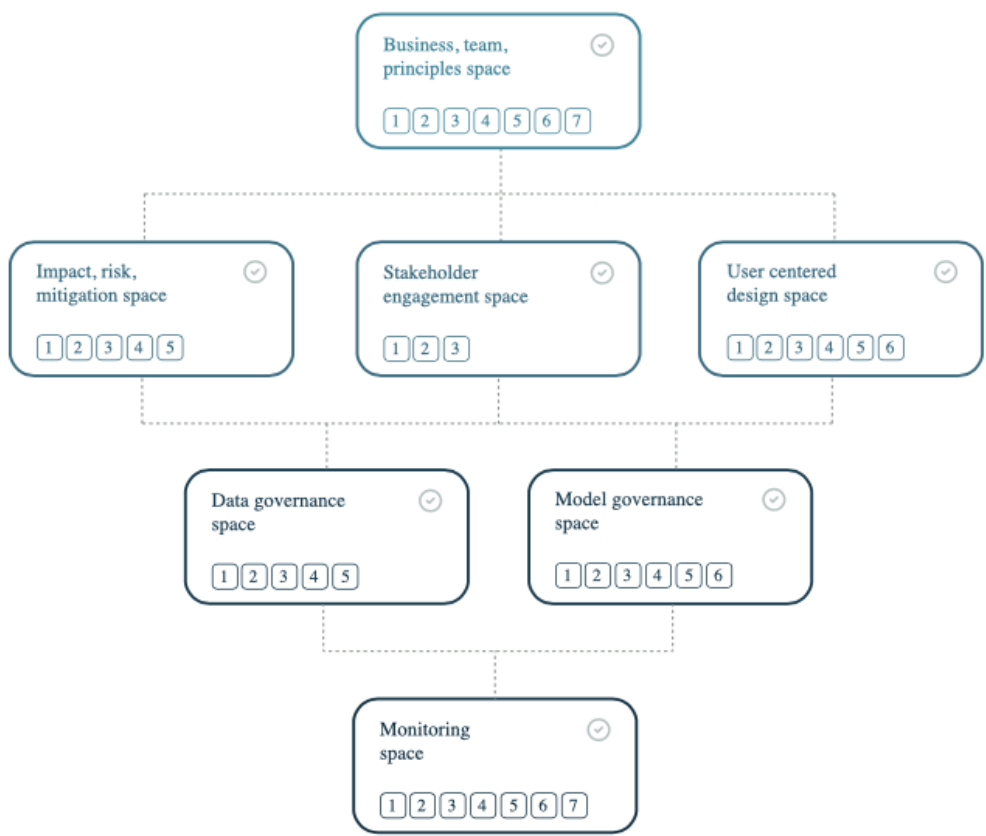

Figure 1. Diagram of the HEAT workflow, representing the interconnected 'spaces'.

## 4.1 Justifying AI Use and Developing Exit Strategies

While the Act assumes building AI systems is a given and mandates compliance, our pro-justice approach requires high-risk AI providers to justify the need of their systems. They are required to demonstrate that building an AI system is not only feasible but also more effective and desirable than alternative solutions considered thus far [35]. Our effort to guide HEAT users through a structured process for making early-stage decisions about whether building an AI system is justified aligns with a feminist AI ethics approach, which challenges the notion and practice of techno-solutionism. Feminist AI ethics advocates for a deeper examination of whether AI is the right tool for the problem at hand, rather than assuming that AI is inherently superior to other solutions. Therefore, in Space 1, the onboarding section of the toolkit, we designed a series of tasks that guide the AI development team through a justification process of their AI system. These tasks include:

- Problem Definition: Define the problem, considering whether it is influenced by inequalities or injustices, and assess how AI might perpetuate or address these issues.
- Role of AI: Describe the functions of AI and explain how these functions contribute to solving the problem.
- Existing Best Practices: Identify current technical and non-technical best practices and their baseline performance. Explain how AI enhances these practices and provides additional value beyond technical metrics.
- Criticisms and Past Incidents: Summarise major critiques and incidents related to AI in similar contexts. Detail how your AI solution addresses these concerns and improves on previous attempts.

This exercise is not just for ethical consideration but also has practical business implications. Addressing the key issue of poor problem formulation and flawed assumptions during the ideation phase [38] —responsible for over 85% of AI innovation project failures [54] — is crucial for business success. When flawed assumptions are embedded within the development and deployment of AI systems, they set the stage for inevitable failure. Regardless of the AI system's accuracy or robustness, it is doomed to fail in its intended purpose if it is assessed against the wrong objectives. This misguided approach guarantees that even the most advanced AI systems will falter, unable to address the true complexities of the problems they were meant to solve. By tackling these foundational problems early, businesses can avoid costly errors.

Moreover, in this exercise, we also emphasise the importance of considering the perspectives of individuals and groups who are directly or indirectly impacted by the AI system. We encourage AI development teams to acknowledge that stakeholders have the legitimate right to reject AI solutions when they are inappropriate. To support this, we provide real-world examples, such as a company's decision to forgo creating a recommendation engine for a gambling product. These examples, collected by AI engineers, illustrate how companies have determined that AI was not the right solution for their needs. These examples of 'exit doors' are intended to empower AI development teams to effectively off-board AI solutions that fail to meet ethical standards or practical needs.

Only after the justification for using AI has been thoroughly completed are users tasked with defining the intended use of the AI system. Documenting the intended use of AI systems is not only a recommended best practice within the responsible AI community, as noted in model cards, but also a mandatory requirement under the Act. However, these practices often assume that the development of an AI system is justified from the outset. In contrast, our approach mandates that AI developers first justify the purpose for AI by clearly defining the problem it aims to solve, explaining why AI is the best solution, and evaluating how it can improve upon existing best practices while addressing any criticisms and incidents related to similar AI applications. It is only after this justification that defining and documenting the intended use of AI becomes meaningful.

4.2 Stakeholder Engagement

While the Act does not make engaging stakeholders external to the development team a compulsory element of the development process, Article 95 specifies that companies can, on a voluntary basis, commit to fulfilling additional requirements, including 'the establishment of inclusive and diverse development teams and the promotion of stakeholders' participation in [the design] process.' As we noted above, the feminist understanding of expertise and the acknowledgment of partiality of knowledge is fundamental to feminist ethics and epistemology and therefore, engaging different stakeholders is not optional in our interpretation of the Act; it is critical to establishing a sound understanding of AI's risks and benefits to specific groups of people (and 'vulnerable' groups in particular, to use the Act's terminology). This is why the users of HEAT are encouraged to return to Space 3 throughout development and deployment, where they are guided through the process of engaging diverse stakeholders. For instance, to consider both intended and unintended consequences of a given AI system for individuals, groups and communities (something that the Act does prescribe) in Space 2, the HEAT users are prompted to move to Space 3 to include marginalised and vulnerable groups in the process of identifying these consequences.

Space 3 provides instructions on how to identify all relevant stakeholders, both direct (end users, decision subjects, system providers and deployers) and indirect (regulators, associated parties, civil society organisations), as well as to distinguish between deployers, users, and affected individuals, as per the Act's categorisation. The central task within Space 3 guides users through engaging stakeholders in a meaningful and power-sensitive way. Drawing on recent critiques of participatory approaches to AI, HEAT highlights the dangers of tokenism and participation-washing, and encourages users to compensate any external stakeholders accordingly for the time they spend interacting with the core design team [8, 19, 46]. While HEAT links users to useful resources, such as the Community Partnerships Playbook [16] and the Design from the Margins report [44], that may help working through the difficulties of participatory projects, Space 3 makes clear that friction is a necessary and unavoidable element of the process. Drawing on a methodology developed by Birhane and colleagues [8], HEAT users are guided through a questionnaire fostering self-reflexivity, including prompts like 'In what ways will the participation process allow for disagreement?' or 'Did participants have the opportunity to refuse participation or withdraw from the process without causing direct or indirect harm to themselves or their communities?'

Following Delgado and colleagues [19], Space 3 presents stakeholder participation projects as ranging from consultation, akin to user experience research which most teams are already conducting in some form, through inclusion, where all stakeholders, not only the development team, have influence over the end product, to collaboration, where stakeholders get to shape the product's scope and purpose through long-term, established partnerships based on mutual trust. While we make clear that the Act does not prescribe such engagements, we suggest that the protection of fundamental rights and values, including human dignity, required by the Act, might necessitate not only consulting diverse stakeholders throughout design but also giving them agency over the nature and outcomes of the design process itself.

We should also note in reference to Space 3 that we are aware of the critiques of the term 'stakeholder', focusing on its roots in extractive and violent colonial contexts [42]. However, since no alternatives have been proposed that would serve as adequate umbrella terms denoting those who have an interest in or are affected – directly and indirectly – by a given project, for ease and clarity (and bearing in mind the needs of users for whom English is a second language), for the purposes of HEAT, we have decided to refer to 'stakeholders.'

4.3 Disability and Design Justice at the Interface Level

Space 4, 'User-centred design' addresses the user-facing elements of an AI-based system. This includes all elements of the interface and interaction design including in relation to usability, access, and transparency. Our pro-justice approach focused on expanding on the Act's minimal requirements in relation to accessibility through the incorporation of disability justice and design justice guidance.

'Disability' is presented at 6 points throughout the Act [14] where it is narrowly framed as a 'vulnerability'. Specifically, disability is mentioned as: 1. a vulnerability to "manipulation" (Recital 29); 2. in relation to appropriate transparency (Recital 132); 3. as part of the prohibition against exploiting vulnerabilities (Article 5); 4. for requiring "appropriate protection" in the context of testing (Article 60) as well as in articles 165 and 95 in the context of voluntary practice.

Furthermore, while the Act no doubt defers much to existing EU and national anti-discrimination and disability rights laws (e.g., the European Accessibility Act [13]), Article 95 explicitly situates testing and the prevention of access harms within 'voluntary codes of conduct': 'assessing and preventing the negative impact of AI systems on vulnerable persons or groups of vulnerable persons, including as regards accessibility for persons with a disability.' This niche and apparently non-mandatory treatment of accessibility seems to belie the reality that 1 in 4 adults in Europe have a disability, while also failing to acknowledge the responsibility of infrastructure developers to create services that are accessible to all.

To remedy this narrow approach, in Space 4 we:
- introduce the social model of disability (emphasising how design can create disability),
- draw attention to 'invisible' disabilities,
- demonstrate how designing for broad access genuinely benefits everyone,
- emphasise the importance of testing with users of diverse abilities
- provide links to relevant law and design justice principles.

In the 'Going further' section we also highlight the representational harms caused by AI for people with disabilities, encourage going beyond just representative testing to deeper involvement and participation, and link to two tools: a Design Justice exercise [1] and the Cards for Humanity tool [12] for toolkit users interested in improving capacity and understanding across their teams.

Of course, disability and design justice principles are incorporated throughout HEAT but the user-facing elements that are the focus of Space 4 present a particularly important and straightforward target for improving practice among AI developers. We want product managers to leave the tool with a broader understanding of disability and awareness of the widespread impact better accessibility design can have for the success of their product as well as for the benefit of society.

4.4 Consent, Complaint & Redress

Consent, complaint and redress are important feminist principles that are expressed to varying extents by the Act. Our tool explores possibilities for making these more pronounced in accordance with adjacent legislation (such as GDPR).

4.4.1 Consent

Article 3 (59) defines consent as "a subject's freely given, specific, unambiguous and voluntary expression of his or her willingness to participate". Although this applies only to participants doing user testing (Article 3, 60, 61; Recital 141 of the AI Act [14]), the GDPR [15] also applies and must be followed in addition to the Act, requiring companies to acquire a data subject's "affirmative" and "unambiguous indication" of consent when processing personal data. However, there is little in the way of guidance on how such consent should best be obtained. We propose a path forward in line with The European Court of Human Rights' pioneering 2003 approach to sexual consent that requires similarly unambiguous and affirmative displays of willingness in order to comply as consent, i.e. not explicitly saying no doesn't disqualify the assault from being viewed as such. Feminist groups campaigned for this more stringent definition because consent is a relatively weak concept, easily strained within relationships characterised by a power imbalance. We view this as analogous to AI company vs. consumer/user tester relationships. Consent as a governance mechanism cannot perform the moral magic of transforming an imbalanced power relationship [21].

Therefore, while our user centred design section (4) includes a variety of questions to consider when obtaining consent before the collection, processing, and storing of personal data, we also include exercises that raise awareness of the difficulty in acquiring consent within such power imbalances. We advocate for participatory design as a method of addressing said imbalances by enabling the direct involvement of affected individuals within the creation of the system. For example, instead of user agreements being created by companies and consented to by affected individuals, as too often is normal practice, agreements could instead be drafted with affected individuals through the methods outlined in stakeholder engagement and user centred design space. In our data governance space, we also ask for a description of how individuals in question consented to the collection of their data, and, where consent was obtained, where consenting individuals were provided with a mechanism to revoke their consent.

4.4.2 Complaint & Redress

The Act does not provide redress mechanisms for individuals but includes: the right to lodge a complaint with a market surveillance authority (Article 85), the right to an explanation for every individual algorithmic decision (Article 86), and the right to report infringements and be protected when reporting (Article 87). It does not guarantee affected persons compensation for harms, and it excludes the private sector from having to mandatorily respect fundamental rights (e.g., the fundamental rights impact assessment in Article 27 is only mandatory for public bodies and private bodies providing public services, excluding big tech companies). However, because fundamental rights are at the heart of its risk-based approach, our tool meets the spirit of the law by guiding companies through creating complaint and redress mechanisms which allow individuals to issue a complaint directly with system providers, acquire a response, and seek redress. We believe this is consistent with the spirit of the law and could also bolster the weaker proposals of: (1) the Product Liability Directive [23], which only covers material harms such as death and destruction to property as a result of a defective AI product, (2) the AI Liability Directive [22], which reduces its own effect by limiting the application to cases in which compensation for fundamental rights is already included in the civil law regulations of the member states (page 9-10 of the AI Liability Directive [22]; see more on this in Wachter [50]  and (3) the GDPR [15], which does require compensation to affected persons only in the narrow scope when harm arises from personal data processed by the algorithms.

In 1.2.1 we suggest that the team appoints a Complaint Oversight Officer to oversee the complaints process. In section 4.5 we include tasks that ensure stakeholders can submit feedback, lodge complaints, or report incidents through establishing non-defensive and constructive Empowerment of Users, Feedback and Reporting, Responsive Communication, Addressing Dissatisfaction, and Acting on Feedback. These serve to integrate complaints into the product for the purpose of improving it, thus responding to feminist work on how, often, "to be heard as complaining is not to be heard" when coming up against an institution, and how the process of complaint (or lack thereof) is indicative of the ethos of the system more broadly [2]. We also establish additional incentives for companies, e.g. that establishing feedback mechanisms not only helps you identify and address stakeholders' concerns but also enhances their experience of and trust in the product, thereby enabling sustainable growth and a competitive advantage.

4.5 How power operates in Data

Increasing user consent in relation to data also involves remedying the power imbalance inherent in data work. While the Act implicitly acknowledges that models trained on incomplete or unrepresentative datasets can produce discriminatory outputs, we aim to address the inadequacy of these implications by emphasising a power-aware perspective [5, 32]. This perspective looks into AI systems but focuses on larger organisational and social contexts, investigating historical inequities, labour conditions, and epistemological standpoints inscribed in data [36]. We emphasise that context-based awareness of these should precede any data collection and processing. Therefore, the first task in our data governance space is Understanding how power operates in data. This informs the way we present real-world examples of the exclusion and underrepresentation of certain groups in datasets [11]. It also informs participatory design practice's emphasis on building trust and fostering collaboration with local communities through transparent communication and involvement in the data collection process [55].

This task drew from a wealth of feminist data science work to debunk those prominent myths in data science: the belief that data is neutral and objective [17], or that harms deriving from data science practices can be remedied by replacing 'bad' data with 'good' data. We ask users to confront how good data is context-specific by asking themselves 'who questions' about their AI systems: who designs it? Who benefits from it? and who is marginalised by it? [20]. Our tool also informs users about the ethical ramifications of data cleaning, tidying, and processing, as well as the selection of training data, data collection methods, data categorization, and the choice of features for analysis [30]. These aspects, often perceived as neutral in machine learning, are business decisions, and therefore are political, strategic and commercial as much as they are technical. Consequently, we stress the importance of documenting the rationale behind decisions and ensuring the meticulous integration of both technical and ethical considerations throughout the process.

4.6 Sustainability and more-than-human-centred design

While the Act [14] acknowledges environmental sustainability of AI systems as an area of concern, it does not mandate providers of high-risk AI to monitor the environmental impact of development. However, in response to the growing prominence of environmental sustainability of AI in the AI ethics discourse [9] – with increased scrutiny on the carbon [48] and water footprints [34] of new models – Article 95 specifies that developers of high-risk AI can voluntarily assess and aim to minimise the negative impact of AI systems on environmental sustainability, focusing on 'energy-efficient programming and techniques for the efficient design, training and use of AI.' Furthermore, Article 112 signals the possibility of introducing additional requirements for high-risk AI providers, including those related to environmental sustainability, when the Act is revised in the future.

Although Articles 95 and 112 do not explicitly require high-risk AI providers to conduct environmental impact assessments and implement mitigation strategies just yet, in our pro-justice interpretation of the Act environmental concerns are seen as inseparable from other concerns about the social impact of AI. Sustainability is not an add-on feature of AI ethics; rather, as human lives are interconnected and interdependent on the lives of other species and environmental well-being in general, care in feminist AI ethics extends beyond 'the human.'

For this reason, Space 2 includes a section dedicated to assessing and mitigating the environmental impact of AI systems. This section introduces users to methods for calculating the carbon footprint of a given model [29] as well as conducting a Life Cycle Analysis for AI systems. It also links users to explainers on approaches such as Green AI [45], which prioritises efficiency in training new models. Beyond these technical and procedural approaches, we have also included, under the 'Go Further' subsection, a speculative design exercise that encourages the users of HEAT to rethink the design of their product at the ideation stage with 'the environment' in mind as an 'unimaginable user persona' [27]. This exercise aims to introduce development teams to more-than-human-centred or allocentric practices in design: approaches to designing that acknowledge the fundamental interconnectedness of all living things, aiming to align human goals with non-human needs and environmental constraints.

## 5. Limitations, Future Directions, and Conclusions

In the practice of compliance with the EU AI Act, the application of pro-justice theories offers a pathway to embody the law's spirit. This pro-justice lens not only ensures that the letter of the law is followed but also that the underlying principles and intentions of the EU AI Act are realized in the development and deployment of AI systems. As highlighted in critiques of the toolkitfication of AI ethics [28], such perspectives are often absent, even with the growing prevalence of responsible AI toolkits. The HEAT toolkit, however, grounds compliance efforts in feminist, anti-racist, and pro-justice critiques, prompting users to engage critically with the power dynamics at play and the structural injustices that shape AI and can be perpetuated by it.

In this paper, we aim to explain how pro-justice theories were applied in designing the toolkit, offering concrete examples to illustrate our approach. However, it is beyond the scope of this paper to describe the full design process, including our collaboration with industry practitioners, the advisory board, and user testing to assess the toolkit's effectiveness and usability. For instance, details on how we ensured the interconnectedness of spaces in the presentation format, as well as the links between content within each space, are not covered here. These aspects will be addressed in future publications.

Looking ahead, future iterations of our tool could allow individual persons to forward their complaints to public interest organisations for them to lodge a collective complaint to the market surveillance authorities. Future versions could also include a resolution process that protects reporting persons from within a company (based on Article 87). Shane Jones [31], whistleblower and former AI Engineering lead at Microsoft complained that "Microsoft does not have a system that I am aware of where you can report a responsible AI issue and have it tracked throughout its lifecycle from the initial report to a resolution". Internal complaint mechanisms can allow companies to avoid PR damage and the authoritarian imposition of restrictions for AI professionals who encounter issues.

Compliance with Ethical Standards

Sources of funding

The project receives funding from the Region of Emilia Romagna, Leverhulme Trust, and Stiftung Mercator.

Disclosure of potential conflicts of interestReferences

1. A Design Justice exercise. https://static1.squarespace.com/static/5d5d34e927fded000105ccc4/t/61f47a25ed03586e68862392/1643412006205/DESIGN%2BJUSTICE%2BZINE_ISSUE1_Activity.pdf Accessed 16 September 202420

https://stpp.fordschool.umich.edu/sites/stpp/files/2024-01/community-partnerships-playbook-2024.pdf

17. Davies, T., Frank, M.: 'There's no such thing as raw data': exploring the socio-technical life of a government dataset. In Proceedings of the 5th Annual ACM Web Science Conference (WebSci '13). Association for Computing Machinery, New York, NY, USA, 75–78 (2013). https://doi.org/10.1145/2464464.2464472

18. Davis, J. L., Williams, A., Yang, M. W.: Algorithmic reparation. Big Data & Society, 8(2) (2021). https://doi.org/10.1177/20539517211044808

19. Delgado, F., Yang, S., Madaio, M., Yang, Q.: The Participatory Turn in AI Design: Theoretical Foundations and the Current State of Practice. Proceedings of the 3rd ACM Conference on Equity and Access in Algorithms, Mechanisms, and Optimization (EAAMO '23). Association for Computing Machinery, New York, NY, USA, Article 37, 1–23 (2023). https://doi.org/10.1145/3617694.3623261

20. D'Ignazio, C., Klein, L.F.: Data Feminism. MIT Press (2023)

21. Edenberg, E., Jones, M. L.: Analyzing the legal roots and moral core of digital consent. New Media & Society, 21(8), 1804-1823 (2019). https://doi.org/10.1177/1461444819831321

22. European Commission: Proposal for a Directive of the European Parliament and of the Council on Adapting Non-contractual Civil Liability Rules to Artificial Intelligence (AI Liability Directive), COM (2022) 496 final (2022). https://eur-lex.europa.eu/legal-content/EN/TXT/?uri=celex%3A52022PC0496

23. European Parliament: European Parliament Legislative Resolution of 12 March 2024 on the Proposal for a Directive of the European Parliament and of the Council on Liability for Defective Products, P9_TA (2024) 0132. https://eur-lex.europa.eu/legal-content/EN/TXT/?uri=EP%3AP9_TA%282024%290132
23

33. Kelly J., Zafar S., Heidemann, L., Zacchi, J., Espinoza, D., Mata, N.: Navigating the EU AI Act: A Methodological Approach to Compliance for Safety-critical Products. 2024 IEEE Conference on Artificial Intelligence (CAI 2024). https://doi.org/10.48550/arXiv.2403.16808

34. Lee, M. S. A., Singh, J.: The Landscape and Gaps in Open Source Fairness Toolkits. In Proceedings of the 2021 CHI Conference on Human Factors in Computing Systems (CHI '21). Association for Computing Machinery, New York, NY, USA, Article 699, 1–13 (2021). https://doi.org/10.1145/3411764.3445261

35. Malgieri, G., Pasquale, F.: Licensing high-risk artificial intelligence: Toward ex ante justification for a disruptive technology, Computer Law & Security Review, Volume 52 (2024). https://doi.org/10.1016/j.clsr.2023.105899

36. Miceli, M., Posada, J., Yang, T.: Studying Up Machine Learning Data: Why Talk About Bias When We Mean Power? Proc. ACM Hum.-Comput. Interact. 6, GROUP, Article 34 (2022). https://doi.org/10.1145/3492853

37. Morley, J., Floridi, L., Kinsey, L., Elhalal, A.: From what to how: an initial review of publicly available AI ethics tools, methods and research to translate principles into practices. Sci. Eng. Ethics 26, 2141–2168 (2020). https://doi.org/10.1007/s11948-019-00165-5

38. Passi, S., Barocas, S.: Problem Formulation and Fairness. In Proceedings of the Conference on Fairness, Accountability, and Transparency (FAT* '19). Association for Computing Machinery, New York, NY, USA, 39–48 (2019). https://doi.org/10.1145/3287560.3287567

39. Peters, D., Loke, L., Ahmadpour, N.: Toolkits, cards and games – a review of analogue tools for collaborative ideation. CoDesign 17(4), 410–434 (2021). https://doi.org/10.1080/15710882.2020.1715444

40. Petterson, A., Cheng, K., Chandra, P.: Playing with Power Tools: Design Toolkits and the Framing of Equity. In Proceedings of the 2023 CHI Conference
25